\documentstyle[11pt,epsfig]{article}

\def\be{\begin{eqnarray}}
\def\ee{\end{eqnarray}}

\def\roughly#1{\mathrel{\raise.3ex\hbox{$#1$\kern-.75em%
\lower1ex\hbox{$\sim$}}}}

\def\gsim{\roughly>}
\def\la{\langle}\def\ra{\rangle}

\def\del{\partial}

\begin{document}

\begin{titlepage}

\hfill {\today }

\begin{center}

\centerline{\Large\bf Dilatons  for Dense Hadronic Matter}

\vspace{.30cm}

Hyun Kyu Lee$^{a}$ and Mannque Rho$^{a,b}$

\vskip 0.50cm

{(a) \it Department of Physics, Hanyang University, Seoul 133-791, Korea}

{(b) \it Institut de Physique Th\'eorique, CEA Saclay, 91191 Gif-sur-Yvette, France}
%{\it 91191 Gif-sur-Yvette, France}

\end{center}
\vskip 1cm

\centerline{\bf Abstract}
\vskip 0.5cm

The idea that the explicit breaking of scale invariance by the trace anomaly of QCD can be rephrased as a spontaneous breaking has been recently exploited to capture the low-energy strong interaction dynamics of dense (and also
hot) matter in terms of two dilaton fields, the ``soft" ($\chi_s$) and the ``hard" ($\chi_h$) fields, in the frame work of the hidden local gauge symmetry. In the  Freund-Nambu model, the spontaneous symmetry breaking of scale symmetry is induced by an explicitly breaking term, while the spontaneous symmetry breaking  is possible  in the flat potential model which is scale symmetric.  We discuss the interplay of the soft and hard dilatons  using the  spontaneously  broken scale symmetry schemes and uncover a novel structure of dense matter hitherto unexplored.

\end{titlepage}\vskip 0.5cm

\section{Introduction}

The emergence of a scale at the quantum level in
a theory that has no scale at the classical level of QCD  has been known as the trace anomaly.
 QCD also has it that chiral symmetry is broken
giving rise to the hadron mass. Thus the chiral symmetry breaking ($\chi$SB) and the scale symmetry
breaking (SSB) are intricately tied to each other.
One can phrase the trace anomaly of QCD
 in terms of a {\em spontaneous breaking}.
Translating the {\em explicit breaking} of scale invariance
into a spontaneous breaking, one can write the trace of the energy-momentum tensor in terms of the
condensate of the dilaton \cite{nambu,Zumino,wetterich,shapozen}. The scale invariance broken spontaneously in that way can then be restored with the dilaton mass going to zero at some transition scale.

In the recent work\cite{leerho} guided by lattice gauge calculations, we suggested the scenario that the ``soft" dilaton ($\chi_s$),
mass goes to zero as
the chiral restoration point is approached, thereby restoring the ``soft" scale symmetry, whereas the
``hard" dilaton ($\chi_h$) stays massive across the chiral transition, with the ``hard" scale symmetry
presumably getting restored at some higher, unidentified (grand-unification?) scale. In \cite{wetterich,shapozen} with a flat potential model, this
scenario is invoked for the hierarchy problem as well as cosmology all the way to the Planck
scale.

The resulting effective Lagrangian for $ SU(2)_L\times SU(2)_R$ chiral symmetry  in terms  of pions, soft and hard dilatons, and the vector mesons ($ \omega_{\mu}$ and $\rho_{\mu})$ as gauge bosons of hidden gauge symmetry, is suggested in the following form,
\be
{\cal L}={\cal L}_{\chi_s}+ {\cal L}_{\chi_h}+{\cal L}_{hWZ}\label{lagtot}
\ee
where
\be
{\cal L}_{\chi_s} &=& \frac{f_\pi^2}{4}\kappa_s^2
%\left(\frac{\chi}{f_\chi}\right)^2
\mbox{Tr}(\partial_\mu U^\dagger \partial^\mu U) + \kappa_s^3 v^3 \mbox{Tr}M(U+U^\dagger)
\nonumber\\
&&
-\frac{f_\pi^2}{4} a\kappa_s^2
%\left(\frac{\chi}{f_\chi}\right)^2
 \mbox{Tr}[\ell_\mu + r_\mu + i(g/2)
( \vec{\tau}\cdot\vec{\rho}_\mu + \omega_\mu)]^2\nonumber\\
&& -\textstyle \frac{1}{4} \displaystyle
\vec{\rho}_{\mu\nu} \cdot \vec{\rho}^{\mu\nu}
-\textstyle \frac{1}{4}  \omega_{\mu\nu} \omega^{\mu\nu}
+\frac 12 \del_\mu\chi_s\del^\mu\chi_s + V(\chi_s)\label{lags}\\
{\cal L}_{\chi_h}&=&\frac 12\del_\mu\chi_h\del^\mu\chi_h +V(\chi_h)\label{lagh}\\
{\cal L}_{hWZ} &=& \textstyle\frac{3}{2} g \left(\frac{\chi_s}{\chi_h}\right)^3 \omega_\mu B^\mu\label{fwzterm}
\ee
where $\kappa_{s}=\chi_{s}/f_{\chi_{s}}$.

There are two features noteworthy with this Lagrangian.

The first is that this Lagrangian treated at the tree order or in the mean field as was done in 1991~\cite{BR91} would give the scaling of the $\rho$ meson of the form
\be
\Phi(n)\equiv \frac{m_\rho^*}{m_\rho}\approx \frac{f_{\chi_s}^*}{f_{\chi_s}}\approx
\frac{f_{\pi}^*}{f_{\pi}}.\label{BRscaling}
\ee
Here the restorations of both symmetries, chiral and scale, coincide.
On the other hand, the HLS Lagrangian (without scalar fields) treated via renormalization group flow would give near the vector manifestation fixed point~\cite{HY:PR} 
\be
\Phi(n) \approx \frac{\la\bar{q}q\ra^*}{\la\bar{q}q\ra}.\label{VMscaling}
\ee

The second feature is that the hWZ term in Eq.(\ref{fwzterm}) plays an important role in the  skyrmions  put on an FCC crystal lattice to simulate dense matter~\cite{PRV-dilaton}. 
The effect of dilatons in the form of Eq.~(\ref{fwzterm}) turns out to be dramatic for both single skyrmion as well as dense skyrmion matter~\cite{PRV-dilaton}. There is a qualitative difference in physics between ``light" dilaton (LD) $m_{\chi} < 4\pi f_\pi\sim 1$ GeV and ``heavy" dilaton (HD) $m_\chi \gsim 1$ GeV.
 The result of the calculation in \cite{PRV-dilaton} shows that  with a HD, the skyrmion matter makes a phase transition at a relatively low density to a half-skyrmion matter with a vanishing quark condensate $\la\bar{q}q\ra^*\rightarrow 0$ but with $f_\pi^*\sim \la\chi\ra^* \neq 0$. This phase corresponds, in HLS theory, to the phase with $a=1$ (i.e., $f_\pi=f_\sigma\neq 0$ in the notation of \cite{HY:PR}) and $g\neq 0$ but $\la\bar{q}q\ra=0$ -- which is unlikely in HLS theory without the dilaton $\chi_s$. This means that in this phase, chiral symmetry is restored but quarks are not deconfined. Since this transition takes place in a theory valid at large $N_c$, it can be identified with what McLerran and Pisarski call ``quarkyonic phase"~\cite{McLerran}. We suggest that this half-skyrmion phase, naturally realized in HLS skyrmions, is generic at high density~\cite{MR:half}.
\section{Two-scalar-field model}
The two scalar model (\ref{fwzterm}) is a special case of generic two-scalar model that can be used to explore the possibility of spontaneous symmetry breaking of the scale symmetry and its physical application of the dilaton field, which emerges as a Goldstone boson.  In fact since 1960's \cite{nambu}\cite{Zumino},  it has been studied in various contexts, including the recent works on cosmology and dark energy problem\cite{shapozen} and dense hadron physics\cite{leerho}. Although long-standing, subtleties still abound in the issue.

The first example is the Freund-Nambu model,  which  has two real scalar fields, $\psi$ and $\phi$, with the potential
 \be
V(\psi, \phi) = V_a + V_b,\label{fn}\ee
where \be V_a &=& \frac{1}{2} f^2 \psi^2 \phi^2 \label{fns} \\
V_b &=& \frac{\tau}{4}[\frac{\phi^2}{g^2} - \frac{1}{2} \phi^4
  - \frac{1}{2g^4}]= \frac{\tau}{8g^4}(g^2\phi^2 -1)^2 \label{fnsb}.
  \ee
$V_a$ is scale invariant and $V_b$ consists of a scale invariant
term $ ~ \phi^4$ and two symmetry breaking terms $\frac{\phi^2}{g^2}$ and $- \frac{1}{2g^4}$.
One can introduce the Goldstone boson, $\chi$, defined by
 \be
\chi = (g^2\phi^2 -1)/(2g),\label{chi}
 \ee
of which the scale transformation
\be
\delta \chi = \epsilon(x \cdot \partial +2) \chi +
\frac{\epsilon}{g},\label{gb}
 \ee
manifests one of the characteristics of Goldstone bosons. In terms of the condensate $\phi_0$ given by the minimum of the potential, $\phi_0=1/g$,
the masses take the form
 \be
 m^2_{\psi} =f^2\phi_0^2, ~~~~ m^2_{\chi} = \tau\phi_0^2.
 \ee
It should be noted that the mass of $\psi$ comes from the scale invariant term $V_a$,  so is independent of $\tau$. Thus it can
have an arbitrary value.  However $m_{\chi}$ depends on $\tau$, going to zero linearly as
$\tau\rightarrow 0$.
This is known to be a characteristic of an approximate
spontaneously-broken scale symmetry~\cite{Zumino}:  Only the mass of one scalar
field (Goldstone boson) must be small, but the masses of all other fields can be arbitrary.

Our study uncovers an interesting possibility: that the local gauge fields of HLS considered in the previous section  can be considered as the $\psi$-type ``matter field" in the Freund-Nambu model. The massive HLS fields  enter (by construction) scale-invariantly, but their masses go to zero as the scale invariance of the soft sector is restored, that is, as $\la\chi_s\ra\rightarrow 0$. This is analogous to that both the mass of the Freund-Nambu dilaton $\phi$ and that of the ``matter field" $\psi$ go to zero as the condensate $\phi_0$ is dialed to zero.

The scale symmetric model in a more general form can be constructed~\cite{wetterich}
\be
{\cal L} = [\partial_{\mu} \chi]^2 + [\partial_{\mu} \phi]^2 -V(\chi, \phi),
\ee
with the potential, V, which is scale invariant in the form
\be
V(\chi, \phi) = v(\phi/\chi) \chi^4. \label{vchi}
\ee
This form can accommodate  the potential Eq.(\ref{fn}) in the Freund-Nambu model as well as the one used in ref. \cite{shapozen}.
Here the scale dimension of both $\la\chi\ra$ and $\la\phi\ra$ is one. Let us consider the potential in which  the vacuum is degenerate in the direction that  $\xi \equiv \phi/\chi $ is constant.  Of course in this form  there is no symmetry breaking term and thus one cannot determine the vacuum expectation values of the scalar fields.  The standard technique is to add a very small symmetry breaking term to fix the vacuum value for a spontaneous symmetry breaking, and take it away as a limiting procedure.

Spontaneous symmetry breaking can be realized by assigning  non-zero vacuum expectation values, $\la\phi\ra_0$ and $\la\chi\ra_0$.
The Goldstone boson field, $\chi$, may be re-parameterized in terms of $\sigma$ as
\be
\chi(x) = Me^{\sigma(x) /M} \label{sigma},
\ee
with the scale transformation
\be
\delta \sigma & \sim & \epsilon(1 +  x_{\nu} \partial^{\nu})\delta + \epsilon M. \label{sigmainf}
 \ee
where the last term, an inhomogeneous term, is typical for a Goldstone boson as in Eq.(\ref{gb}). The Higgs field $\phi$ gets mass~\cite{shapozen}
\be
m^2_{\phi} = 4 \lambda \chi^2_0 \xi^2
\ee
where $\tilde{\phi}_0 = \xi M$ has been used.  In this scheme, where the scale symmetry is protected at quantum level~\cite{shapozen},   the mass scale emerges, however, from the (spontaneous) symmetry-breaking parameters figuring in the Higgs mass.

In the scheme where the mass scale emerges from the scale-invariant QCD due to spontaneous symmetry breaking of the scale symmetry,  it is natural that a Goldstone boson appears in the effective theory.   In the previous section, we introduced two dilatons as Goldstone bosons, the condensates of which are supposed to have  different  temperature or density dependencies.   We may suppose these to be different ways of realizing the SSB of scale symmetry.  For example, $\chi_s$ can be identified with $\sigma$ in Eq.(\ref{sigma}) and $\chi_h$ with $\chi$ in Eq.(\ref{chi}), such that most of the bag constant can be explained by the Freund-Nambu type of the model where the spontaneous symmetry breaking is induced by the explicit symmetry breaking term in the Lagrangian.  Then  $\la\sigma\ra$ from the scale-invariant effective potential at quantum level melts down at a lower temperature as the lattice calculations are suggesting.  It is an interesting question whether the origin of the different temperature/density dependence of the vacuum expectation values of two dilatons is due to the different manifestations of SSB of scale symmetry. We hope to address this issue in a future publication.

\section{Discussion}

Using  a Lagrangian with an explicit breaking term put in the (effective)
Lagrangian which is transformed into a spontaneous breaking, it is possible to  deal with the situation in which the scale symmetry is broken both
explicitly and spontaneously. In this way we are able to connect $\chi$SB which generates masses to
scale symmetry breaking which also generates masses. This would
allow us to connect the spontaneously broken chiral symmetry  to other condensates such as kaon condensate.
By identifying the two-component structure of the QCD trace anomaly with a ``soft" dilaton and a
``hard" dilaton, it is found to be possible to make the connection between the scaling of hadron properties
based on Brown-Rho scaling -- which was originally anchored on spontaneous breaking of scale invariance in
medium -- and that based on renormalization group flow of hidden local symmetry in the vector manifestation. For this the scalar degree of freedom plays a crucial role. It is highly nontrivial that the vanishing of the vector meson mass at the vector manifestation fixed point in HLS can be related to
the dilaton condensate going to zero representing a spontaneously broken scale symmetry being
{\em restored}.  It is found that the two component dilaton structure which renders the
scaling behavior of the homogeneous Wess-Zumino term of the
$\omega\cdot B$-type consistent with scale invariance of the Lagrangian plays an extremely
important role in the structure of the nucleon, nuclear and dense
matter as found in \cite{PRV-dilaton}.

\subsection*{Acknowledgments}
We are grateful for discussions with Gerry Brown, Masa Harada, Byung-Yoon Park and Vicente Vento for discussions. This work is supported by the WCU project of Korean Ministry of Education, Science and Technology (R33-2008-000-10087-0).

\end{document}